# MULTI-VANADIUM SUBSTITUTED POLYOXOMETALATES AS EFFICIENT ELECTROCATALYSTS FOR THE OXIDATION OF L-CYSTEINE AT LOW POTENTIAL ON GLASSY CARBON ELECTRODES

**Bineta Keita[a], Roland Contant[b], Pierre Mialane[c], Francis Sécheresse[c], Pedro de Oliveira[a], Louis Nadjo[a]\***

[a]Laboratoire de Chimie Physique, Groupe d'Electrochimie et de Photoélectrochime,
UMR 8000, CNRS, Université Paris-Sud, Bâtiment 350, 91405 Orsay Cedex, France

[b]Retired

[c]Institut Lavoisier, UMR 8180, Université de Versailles St Quentin
45, avenue des Etats-Unis, 78035 Versailles

tel: 33 1 69 15 77 51
fax: 33 1 69 15 43 28
e mail: nadjo@lcp.u-psud.fr

**Abstract**

In search for efficient electrocatalysts for the oxidation of L-cysteine on glassy carbon, multi-vanadium-substituted polyoxometalates (POMs, for short) were tested. The electrochemical behaviours of the sandwich-type complex $[As_2W_{18}(VO)_3O_{66}]^{11-}$ were studied in a pH 7 medium and compared with those of the three following Dawson-type vanadium-substituted complexes: $[P_2V_2W_{16}O_{62}]^{8-}$ ($P_2V_2W_{16}$), $[P_2MoV_2W_{15}O_{62}]^{8-}$ ($P_2MoV_2W_{15}$) and $[P_2V_3W_{15}O_{62}]^{9-}$ ($P_2V_3W_{15}$). Electrochemistry shows that the sandwich-type POM contains 2 $V^{IV}$ centers and one $V^V$ center and must be formulated $As_2V_2^{IV}VW_{18}$, in agreement with titration, elemental analysis and magnetic measurements on this element. The two-electron composite wave that features the redox behaviour of the two $V^{IV}$ centers of $As_2V_2^{IV}VW_{18}$ are very close to each other and remains practically merged whatever the potential scan rate. In addition of this sharp contrast with the behaviours of the two $V^V$ centers of $P_2V_2W_{16}$ or $P_2MoV_2W_{15}$, the very slow electron transfer kinetics associated with the second wave of $P_2V_3W_{15}$ appears also at variance with the case of $As_2V_2^{IV}VW_{18}$. All the POMs of this work proved efficient for the oxidation of L-cysteine. Comparison of the present results with those of mono-Vanadium substituted POMs indicates that accumulation of vanadium atoms in the POM framework is beneficial in the electrocatalytic process. In addition, the present work highlights the important influence of the POM structure in the electrocatalytic oxidation of L-cysteine. The remarkable outcome of this work is that the potential for the oxidation of L-cysteine in the presence of the selected POMs has been substantially driven in the negative direction compared to the case of glassy carbon alone, a feature which is associated with faster kinetics. The stability of the systems must also be pointed out.

## 1. Introduction

L-cysteine is recognized as one of the most important among marker thiols in biological systems [1-4]. It is used in medicine and food industries. As a consequence, its oxidation and accurate determination at low potential constitute a valuable task and the search for sensitive and selective methods in these purposes continues unabated [1-4]. In this context, electrochemical techniques at solid electrodes appear among the favourites, probably because, simultaneously with detection, they provide interesting insights into the redox behaviours of cysteine. Unfortunately, direct oxidation of thiols including cysteine, at classical solid electrodes surfaces (Pt, Au, Graphite, Carbon) is kinetically slow and requires a large overpotential to be effective. Hence the necessity to study electrocatalytic processes that might decrease the overpotential [1-13]. Also associated with this strategy are several other beneficial improvements including the increase in selectivity, the decrease of surface fouling and the decrease of surface oxide formation. Remarkable achievements were obtained recently with boron-doped diamond electrodes [2,7,11] and also with carbon nanotube materials, unmodified or modified with various catalysts [1-3,9].

We now report multi-vanadium-substituted polyoxometalates (POMs) which act as mediators to improve the electro-oxidation of cysteine on glassy carbon electrodes (GC). Bare carbon electrodes, including bare glassy carbon, show a very slow kinetics for this process [4,14]. To our knowledge, as concerns POMs, only $[Co^{III}W_{12}O_{40}]^{5-}$ was used to oxidize cysteine in aqueous acidic media [15], but the formal potential of the $Co^{III}/Co^{II}$ couple within this molecule is more positive than + 0.7 V vs SCE. All the selected vanadium-substituted POMs are stable in a large pH domain from 0 to 8; furthermore, solutions of their oxidized ($V^V$) and reduced ($V^{IV}$) forms are stable in the open air. These classical complexes of the Dawson series ($[P_2V_2W_{16}O_{62}]^{8-}$ ($P_2V_2W_{16}$) and $[P_2MoV_2W_{15}O_{62}]^{8-}$ ($P_2MoV_2W_{15}$) and $[P_2V_3W_{15}O_{62}]^{9-}$ ($P_2V_3W_{15}$)) proved efficient previously in the electrocatalytic reduction of NO and the oxidation of coenzymes like NAD(P)H [16]. They are sufficiently powerful oxidants to oxidize L-cysteine. Additionally, one sandwich-type POM was used, of formula $[As_2W_{18}(VO)_3O_{66}]^{11-}$ ($As_2V_3W_{18}$) [17]. The formal potentials of all these POMs are lower by 0.3 to 0.4 V than that of the $Co^{III}/Co^{II}$ couple within $[Co^{III}W_{12}O_{40}]^{5-}$.

## 2. Experimental

**Synthesis** All the phosphotungstates in the present study are primarily derived from a-$[P_2W_{18}O_{62}]^{6-}$ and were synthesized by published methods [16]. $K_{11}[As_2W_{18}(VO)_3O_{66}] \cdot 23\,H_2O$ was also obtained by the published method [17].

**General Methods and Materials.** Pure water was used throughout. It was obtained by passing through a RiOs 8 unit followed by a Millipore-Q Academic purification set. All reagents were of high-purity grade and were used as purchased without further purification. The UV-visible spectra were recorded on a Perkin-Elmer Lambda 19 spectrophotometer on $1.6 \times 10^{-5}$ M solutions of the relevant polyanion. Matched 1.000 cm optical path quartz cuvettes were used. The compositions of the various media were as follows: for pH 4 and 5: 1 M $CH_3COONa + CH_3COOH$ ; for pH 7: 1 M $NaCl + 50$ mM $NaH_2PO_4 + NaOH$; or 0.5 M $Li_2SO_4 + 0.2$ M Tris $+ H_2SO_4$.

**Electrochemical Experiments.** The same media as for UV-visible spectroscopy were used for electrochemistry, but the polyanion concentration was $2 \times 10^{-4}$ M.
The solutions were thoroughly deaerated with pure argon for at least 30 min. and kept under a positive pressure of this gas during the experiments. The source, mounting and polishing of the glassy carbon (GC, Tokai, Japan) electrodes has been described [18]. The glassy carbon samples had a diameter of 3 mm. The electrochemical set-up was an EG & G 273 A driven by a PC with the M270 software. Potentials are quoted against a saturated calomel electrode (SCE). The counter electrode was a platinum gauze of large surface area. All experiments were performed at room temperature.

Upon addition of a small amount of cysteine to one of the selected POMs with its V-center(s) in the oxidised form, the colour associated with $V^{IV}$ develops immediately, thus indicating the reduction of the POM. The appropriate amount of cysteine was added until complete reduction of the POM, as followed by UV-visible spectroscopy. Such a procedure is interesting in that it provides a stoichiometry for the cysteine/POM reaction and indicates simultaneously the reduced POM to remain stable in the reaction medium. It was also checked that this reduction could be obtained by controlled potential electrolysis of the POM alone. In the following, electrocatalysis experiments will be started with the relevant V-center(s) in the reduced state. AN analogous procedure was used previously in the study of the electrocatalytic oxidation of NADH by selected POMs [19].

## 3. Results and Discussion

### 3.1. Electrocatalysis of L-cysteine oxidation by multi-vanadium substituted Dawson-type POMs.

The electrochemistry of the three Dawson-type POMs considered in this work was described previously [16]. For all these compounds, it was checked that the vanadium sites are the only active centers for the catalysis of L-cysteine oxidation. As a consequence, interest is focused only on these substituents. It is worth reminding briefly the main behaviours of these POMs. The redox processes of their V-centers were observed in a potential domain well positive of those of the Mo- and W-centers [16]. $P_2V_3W_{15}$ and $P_2V_2W_{16}$ are stable between pH 0 and 8; the presence of the Mo atom narrows somewhat this domain for $P_2MoV_2W_{15}$ which is stable from pH 0 to pH 7. Whatever the pH in these stability domains, separate redox processes were observed for V-centers. Their oxidising power, as judged from their formal potentials, is far lower for the trivanadic species compared to those of the divanadic POMs, which have close formal potentials. For example, at pH 7, these formal potentials are 122 mV, 298 mV and 309 mV vs SCE respectively for $P_2V_3W_{15}$, $P_2V_2W_{16}$ and $P_2MoV_2W_{15}$.

For clarity in the forthcoming descriptions, Figure 1A shows the cyclic voltammogram (CV) associated with the V-centers of $P_2V_2W_{16}$: it is constituted by a first one-electron reversible wave followed by a composite wave, largely irreversible chemically. Prior to electrocatalysis experiments, the POMs are in their reduced state as explained in the Experimental Section. Upon addition of L-cysteine in the medium, a current increase is observed readily at the oxidation potential of the first V-wave. The high efficiency of the electrocatalysis is worth noting. In the potential domain where this catalysis is observed, the direct oxidation of L-cysteine on the glassy carbon electrode surface is negligible [1,20]. The catalytic activity was observed to increase with pH; even so, it remains significant at pH 5 as appears from Figure 2. At this pH where the optimum catalytic activity is not obtained, the Figure indicates that the current increases at the potential of the first V-wave even for small values of the excess parameter $\gamma$ ($\gamma = C°_{L\text{-cysteine}} / C°_{POM}$). Also, it must be pointed out that the electrocatalysis is observed in the same potential domain as in the pH 7 medium, because the characteristics of the first V-redox couple does no longer vary with pH for pH values > 4 [16].

In the following, the catalytic efficiencies of the other POMs will be compared with that of $P_2V_2W_{16}$ in the pH 7 buffer, a medium useful for the study of biological systems.

Figure 3A shows the catalytic activities, for the oxidation of L-cysteine, to be comparable for the two divanadic species, $P_2V_2W_{16}$ and $P_2MoV_2W_{15}$. However, due to its higher stability, $P_2V_2W_{16}$ is a better candidate for long term applications in the pH 7 to 8 media suitable for biological systems. The tri-V-substituted POM, $P_2V_3W_{15}$ is also very stable in these media. However, Figure 3B shows its catalytic activity to be much lower than that of $P_2V_2W_{16}$. Indeed, this Figure does show that the presence of $P_2V_3W_{15}$ makes the oxidation of L-cysteine to occur in a potential domain more negative than in its absence. Nonetheless, the CVs indicate an overall electron transfer kinetics during the catalysis much slower than in the presence of $P_2V_2W_{16}$. Tentatively, the difference in formal potentials of these two catalysts might explain this observation (122 mV and 298 mV for $P_2V_3W_{15}$ and $P_2V_2W_{16}$, respectively).

**3.2. Comparison, in the pH 7 medium, of the electrochemical and electrocatalytic properties of the multi-vanadium-substituted Dawson-type POMs and the sandwich-type POM $[As_2W_{18}(VO)_3O_{66}]^{11-}$ ($As_2V_3W_{18}$)**

### 3.2.1. Electrochemistry

Actually, the POM $As_2V_3W_{18}$ must be formulated $As_2V_2^{IV}VW_{18}$ because it is a mixed valence sandwich-type species with the vanadium atoms in two different oxidation states ($V^V$ and $V^{IV}$). To our knowledge, the electrochemistry of this POM was not described previously. For completeness of the description of its CV, it is pointed out that the reduction of the tungsten centers appear at – 1.130 V vs SCE, close to the supporting electrolyte discharge. However, as stated in the preceding section, the present study focuses on the redox behaviours of V-centers, which constitute the active element for the oxidation of L-cysteine. Figure 4A represents the CV associated, at pH 7, with the $V_2^{IV}V$ centers within $As_2V_2^{IV}VW_{18}$. Worth of notice, the redox processes of the V-centers are located in a potential domain (roughly - 0.400 V to + 0.600 V) well positive of that of W-centers (- 0.800 V to – 1.130 V). In other words, the vanadium centers are much more easily reduced than the W-centers, in agreement with previous observations with mono- or multi-V-substituted POMs [16,21,22]. The CV of $V_2^{IV}V$ centers in Figure 4A shows a reversible reduction wave located at – 0.220 V and a large current intensity composite and reversible oxidation wave, the main peak of which is located at + 0.296 V. These waves are attributed respectively to the one electron reduction of the $V^V$ center and to the two-electron oxidation of the two $V^{IV}$ centers. For clarity, it is worth emphasizing that the two $V^{IV}$ centers are associated with two, very closely spaced, chemically reversible waves, which can, however, be distinguished on scrutiny of Figure 4A. Controlled potential coulometry performed at + 0.400 V confirms that two electrons per molecule are actually consumed for the whole oxidation process. During this electrolysis, a colour change is also observed from the brown colour of $V^{IV}$ centers to the yellow colour of $V^V$ centers. In short, electrochemistry shows $As_2V_2^{IV}VW_{18}$ to contain 2 $V^{IV}$ centers and one $V^V$ center, in agreement with titration, elemental analysis and magnetic measurements on this element.

The results obtained at pH 7 indicate that the characteristics of the redox processes of $As_2V_2^{IV}VW_{18}$ are very different from those of the three Dawson-type V-substituted POMs studied in the preceding section. The most important difference concerns the redox behaviour of the two $V^{IV}$ centers of $As_2V_2^{IV}VW_{18}$. The two-electron composite wave that features this process shows that the two anticipated one-electron processes are very close from each other. It is worth noting that these waves remain practically merged whatever the potential scan rate in the range from 2 mV s$^{-1}$ to 500 mV s$^{-1}$. Indeed, a sharp contrast exists with the behaviours of the two $V^V$ centers of $P_2V_2W_{16}$ or $P_2MoV_2W_{15}$ [16]. It must be reminded that Figure 1A illustrates the example of $P_2V_2W_{16}$ in which the two one-electron steps are characterised by two waves with their reduction peaks separated by 0.596 V and the reoxidation peaks separated by 0.250 V. The electron transfer in the second redox couple of $P_2MoV_2W_{15}$ is faster than observed for the corresponding process of $P_2V_2W_{16}$. However, the two redox couples of $P_2MoV_2W_{15}$ remain distinct, and the corresponding reduction peaks and reoxidation peaks are separated respectively by 0.514 V and 0.308 V. In contrast, a loose similarity can be found between $P_2V_3W_{15}$ and $As_2V_2^{IV}VW_{18}$: in both POMs, the redox processes are featured by a first reversible wave, followed by a composite wave, the two steps of which are practically merged [16]. However, at variance with the case of $As_2V_2^{IV}VW_{18}$, very slow electron transfer kinetics are associated with this second wave in the case of $P_2V_3W_{15}$. Finally, the differences observed between the sandwich-type POM and the three Dawson type POMs might, likely, be explained by the differences between their overall acido-basic properties.

In summary, both in $P_2V_3W_{15}$ and $As_2V_2^{IV}VW_{18}$, two of their V-centers are featured by a two-electron composite wave, in contrast with the case of the di-vanadium substituted Dawson type POMs which show two very separated one-electron waves.

### 3.2.2. Electrocatalysis

Figure 4B indicates that the electrocatalytic oxidation of L-cysteine begins readily on the two-closely spaced oxidation waves of the two $V^{IV}$ centers of $As_2V_2^{IV}VW_{18}$. Here again, two observations must be stressed: i) the two V-centers of interest are in the reduced state prior to the beginning of the electrocatalysis experiments; ii) the redox processes associated with the two $V^{IV}$ centers are chemically reversAible, at least in the timescale of slow cyclic voltammetry. Qualitatively, the large current intensity of this catalytic wave and its chemical irreversibility indicate the efficiency of the reaction, as explained previously in the case of di-vanadium substituted Dawson-type POMs. However, with $As_2V_2^{IV}VW_{18}$, the catalysis occurs with a substantial potential gain which amounts to 0.120 V and 0.150 V compared respectively to the cases of $P_2V_2W_{16}$ or $P_2MoV_2W_{15}$. It must be kept in mind that, with Dawson-type POMs, the peak of the catalytic process is located also close to their respective first one-electron waves, active for the electrocatalytic oxidation of L-cysteine. However, these waves are located at more positive potentials than the composite two-electron wave of $As_2V_2^{IV}VW_{18}$ which triggers the same catalytic process.

Finally, among the POMs studied in this work, $As_2V_2^{IV}VW_{18}$ seems to be the best candidate, besides $P_2V_2W_{16}$, for the catalytic oxidation of L-cysteine. Furthermore, no deactivation of the electrode is observed during several potential cycles, a behaviour worth of notice. In addition, an interval of the excess parameter $\gamma = 1$ to at least $\gamma = 20$ was tested and the catalytic current was found to vary linearly in this domain. At present, several hour stability has been observed for the reduced POM in the presence of cystein or cystine by UV-visible spectroscopy. The study of the long-term stability of the electrocatalytic system after thousands cycles is beyond the scope of this paper.

It was found interesting to compare the activity of $As_2V_2^{IV}VW_{18}$ with that of $H_4PVW_{17}$ which has the less positive formal potential among the mono-vanadium-substituted Dawson-type POMs. For this purpose, double potential step chronocoulometry [19] was used to determine the second order rate constant of the reaction with the sandwich-type complex as performed previously with the mono-substituted complex [20]. The presence of an excess $\gamma = 10$ of L-cysteine was used to establish pseudo first order conditions. The potential step programs were from 0 V to + 0.5 V and back to 0 V for $As_2V_2^{IV}VW_{18}$ and from - 0.2 V to + 0.5 V and back to – 0.2 V for $H_4PVW_{17}$. The results indicate that $As_2V_2^{IV}VW_{18}$ is roughly five times more efficient than $H_4PVW_{17}$, as judged from the rate constants k = 5025 $M^{-1}$ $s^{-1}$ and 1000 $M^{-1}$ $s^{-1}$ for the two complexes respectively. The increased number of vanadium atoms appears to be beneficial to the electrocatalytic process. The relatively high rate constants indicate substantial improvement of the kinetics of L-cysteine oxidation on the glassy carbon electrode surface. This feature must be constrasted with the known slowness and sluggishness of the oxidation of thiols at solid electrodes, including glassy carbon, on which the reaction requires at least 1.0 V to be effective [1,7].

All the results together suggest that the structure of the POM intervenes as an important parameter : as a matter of fact, $P_2V_3W_{15}$ is the weaker electrocatalyst among all the POMs studied in this work, even though it contains the same number of V-atoms as $As_2V_2^{IV}VW_{18}$. Accumulation of vanadium atoms in the POM framework is beneficial as was also observed previously for several sandwich-type POMs in the electrocatalytic reduction of the NOx [23-27] or that of $O_2$ [28-30]. In addition, the present work highlights the important influence of the POM structure in the electrocatalytic oxidation of L-cysteine. Finally, the two trivanadic POMs show an overall two-electron redox process for two of their V-centers; however, these waves, in the case of $P_2V_3W_{15}$, are located in a too negative potential domain (between – 0.050 V and – 0.660 V) for an efficient electrocatalytic oxidation of L-cysteine.

**Conclusion and future work**

The remarkable outcome of this work is that the potential for the oxidation of L-cysteine in the presence of the selected POMs has been substantially driven in the negative direction compared to the case of glassy carbon alone, a feature that stresses the importance of the electrocatalysis. The stability of the systems must also be pointed out.

Comparison of the electrochemistry of several V-containing POMs and study of their electrocatalytic properties toward the oxidation of L-cysteine highlight behaviours that might be ultimately traced to their respective structures. The sandwich-type complex shows promise among these complexes. Owing to the overall bielectronic process associated with two of its vanadium centers, $As_2V_2^{IV}VW_{18}$ is probably a good candidate for numerous electrocatalytic processes that need two electrons to proceed. Furthermore, its stability and the absence of inhibition phenomena during potential cycling experiments entitle us to envision it for the fabrication of chemically modified electrodes for analytical purposes. Work is progress in this direction.

The important decrease of the overpotential for cysteine oxidation raises the question of applicability of the systems in complex biological media, where other analytes might interfere. Preliminary experiments with $As_2V_2^{IV}VW_{18}$ and glutathione show promise for an important selectivity in favour of cysteine. Work in progress will also examine other relevant biological species.

**Acknowledgement** This work was supported by the CNRS (UMR 8000 and UMR 8180), Université Paris-Sud XI and Université de Versailles St Quentin.

**References**

[1] H. Han, H. Tachikawa Frontiers in Bioscience 10 (2005) 931.

[2] O. Nekrassova, N.S. Lawrence, R.G. Compton Electroanalysis, 16 (2004) 1285.

[3] A. Salimi, R. Hallaj Talanta 66 (2005) 967.

[4] R.D. Thackrey, T.L. Riechel J. Electroanal. Chem. 245 (1988) 131.

[5] Z-N. Gao, J. Zhang, W-Y. Liu J. Electroanal. Chem. 580 (2005) 9.

[6] S. Cakir, E. Bicer Bioelectrochemistry 64 (2004) 1.

[7] C. Terashima, T.N. Rao, B.V. Sarada, Y. Kubota, A. Fujishima Anal. Chem. 75 (2003) 1564.

[8] S. Fei, J. Chen, S. Yao, G. Deng, D. He, Y. Kuang Analytical Biochemistry 339 (2005), 29.

[9] Y-D. Zhao, W-D. Zhang, H. Chen, Q-M Luo Sensors and Actuators B 92 (2003) 279.

[10] D. Mimica, F. Bedioui, J.H. Zagal Electrochimica Acta 48 (2002) 323.

[11] N. Spataru, B.V. Sarada, E. Popa, D.A.Tryk, A. Fujishima Anal. Chem. 73 (2001) 514-5190.

[12] K. Ozoemena, P. Westbroek, T. Nyokong Biochemistry Communications 3 (20001) 529
.
[13] S. Maree, T. Nyokong J. Electroanal.Chem. 492 (2000) 120.

[14] J.A. Reynaud, B. Maltoy, P. Canessan J. Electroanal. Chem. 114 (1980) 195.

[15] G.A. Ayoko, M.A. Olatunji Polyhedron 2 (1983) 577.

[16] B. Keita, I.M. Mbomekalle, L. Nadjo, P. de Oliveira, A. Ranjbari, R. Contant, C.R. Chimie 8 (2005) 1057.

[17] P. Mialane, J. Marrot, E. Rivière, J. Nebout, G. Hervé Inorg. Chem. 40 (2001) 44.

[18] B. Keita, F. Girard, L. Nadjo, R. Contant, R. Belghiche, M. Abbessi, J. Electroanal. Chem. 508 (2001) 70.

[19] B. Keita, K. Essaadi, L. Nadjo, R. Contant, and Y. Justum (1996). *J. Electroanal. Chem.* 404 (1996) 271.

[20] B. Keita, I.M. Mbomekalle, P. de Oliveira, A. Ranjbari, Y. Justum ; L. Nadjo, D. Pompon J. Cluster Science, in press, 2006.

[21] B. Keita, I.M. Mbomekalle, L. Nadjo, C. Haut Electrochem. Commun. 6 (2004) 978-983.

[22] E. Cadot, M. Fournier, A. Tézé, G. Hervé Inorg. Chem. 35 (1996) 282.

[23] I.M. Mbomekalle, R. Cao, K.I. Hardcastle, C.L. Hill, M. Ammam, B. Keita and L. Nadjo, T.M. Anderson C.R. 8 (2005) 1077.

[24] B. Keita, I.M. Mbomekalle, L. Nadjo, R. Contant *Electrochem. Commun.* 2001, 3, 267.

[25] B. Keita, I.M. Mbomekalle, L. Nadjo *Electrochem. Commun.,* 2003, 5, 830.

[26] D. Jabbour, B. Keita, I.M. Mbomekalle, L. Nadjo, U. Kortz, Eur. J. Inorg. Chem. (2004) 2036.

[27] L-H. Bi, U. Kortz, S. Nellutla, A.C. Stowe, N.S. Dalal, B. Keita, L.Nadjo Inorg. Chem. 44 (2005) 896.

[28]   I.M. Mbomekalle, B. Keita, L. Nadjo, P. Berthet, K.I. Hardcastle, C.L. Hill, T.M. Anderson Inorg. Chem. 42 (2003) 1163.

[29]   I.M. Mbomekalle, B. Keita, L. Nadjo, W.A. Neiwert, L. Zhang, K.I. Hardcastle, C.L. Hill, T.M. Anderson Eur. J. Inorg. Chem. (2003) 3924.

[30]   B. Keita, I.M. Mbomekalle, Y.W. Lu, L. Nadjo, P. Berthet, T.M. Anderson, C.L. Hill *Eur. J. Inorg. Chem.* (2004) 3462.

**Figure Caption**

**Figure 1.** Cyclic voltammograms (CVs) of $2\times10^{-4}$ M $P_2W_{16}V_2$ restricted to V-centred redox processes in a pH 7 medium. The scan rate was 2 mV s$^{-1}$, the working electrode was glassy carbon, and the reference electrode was SCE.
  **A.** pattern showing the two V-waves
  B. CV showing the two V-waves in the presence of various quantities of L-cysteine. The excess parameter γ is defined as: $\gamma = C°_{\text{L-cysteine}} / C°_{\text{POM}}$.
For further details, see text.

**Figure 2.** Cyclic voltammograms (CVs) of $2\times10^{-4}$ M $P_2W_{16}V_2$ restricted to V-centred redox processes in a pH 5 medium, in the presence of various quantities of L-cysteine. The excess parameter γ is defined as: $\gamma = C°_{\text{L-cysteine}} / C°_{\text{POM}}$. The scan rate was 2 mV s$^{-1}$, the working electrode was glassy carbon, and the reference electrode was SCE.
For further details, see text.

**Figure 3.** Cyclic voltammograms (CVs) of $2\times10^{-4}$ M V-substituted Dawson-type POMs restricted to V-centred redox processes in a pH 7 medium in the presence of an excess γ = 10 of L-cysteine. The excess parameter γ is defined as: $\gamma = C°_{\text{L-cysteine}} / C°_{\text{POM}}$. The scan rate was 2 mV s$^{-1}$, the working electrode was glassy carbon, and the reference electrode was SCE.
  **A.** the V-substituted POMs are $P_2V_2W_{16}$ and $P_2MoV_2W_{15}$.
  **B.** the V-substituted POMs are $P_2V_2W_{16}$ and $P_2V_3W_{15}$. The same amount of L-cysteine present in the POM solutions was also added to the electrolyte.
For further details, see text.

**Figure 4.** Cyclic voltammograms (CVs) of $2\times10^{-4}$ M $As_2V_2^{IV}VW_{18}$ restricted to V-centred redox processes in a pH 7 medium. The scan rate was 2 mV s$^{-1}$, the working electrode was glassy carbon, and the reference electrode was SCE.
  **A.** pattern showing the two V-waves
  B. CV showing the two V-waves in the presence of an excess γ = 10 of L-cysteine.. The excess parameter γ is defined as: $\gamma = C°_{\text{L-cysteine}} / C°_{\text{POM}}$.
For further details, see text.

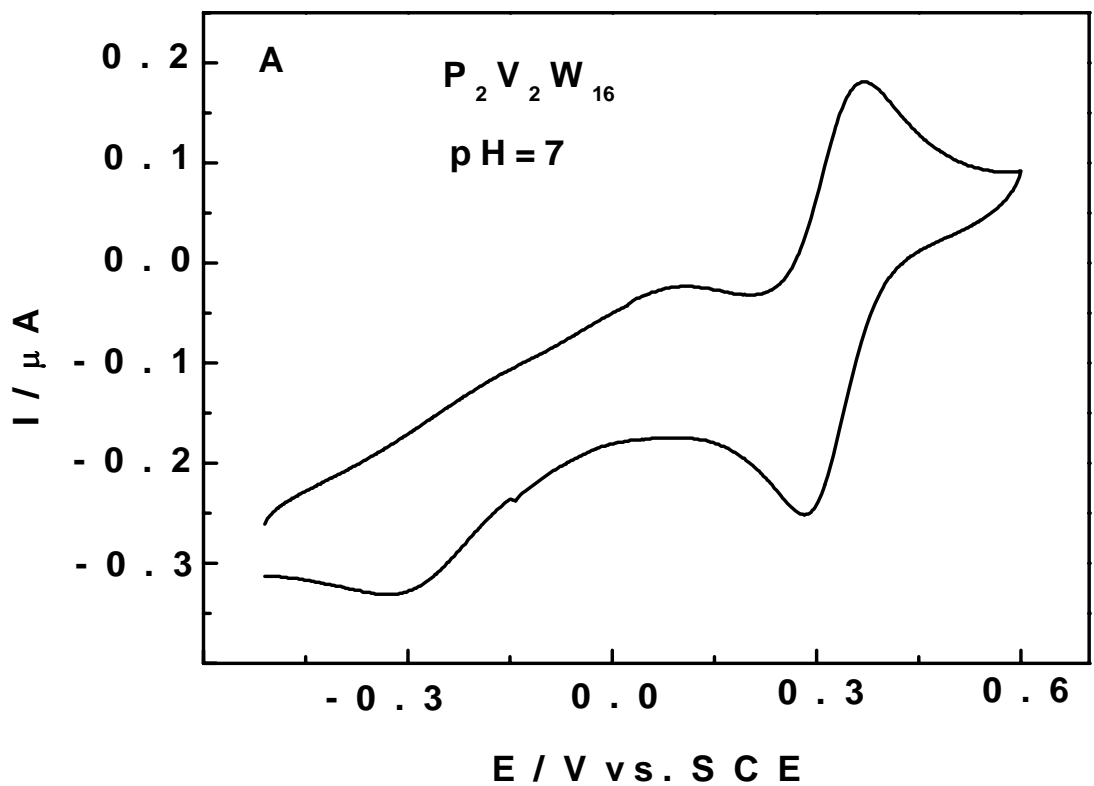

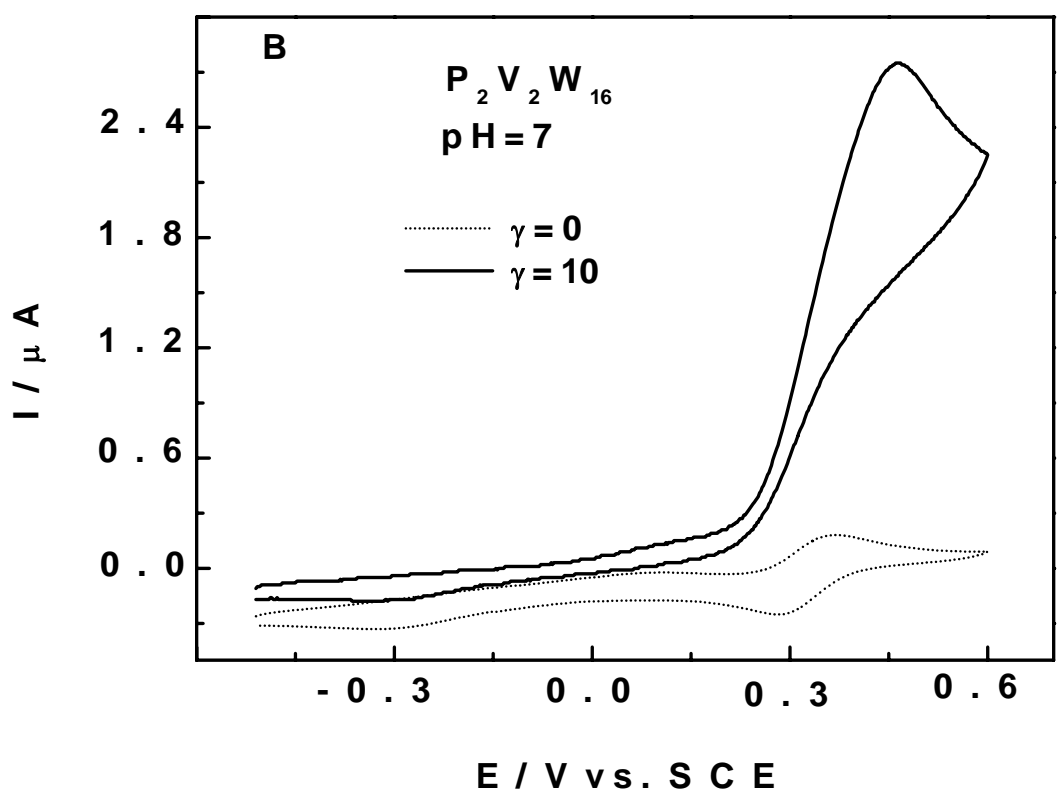

**Figure 1**

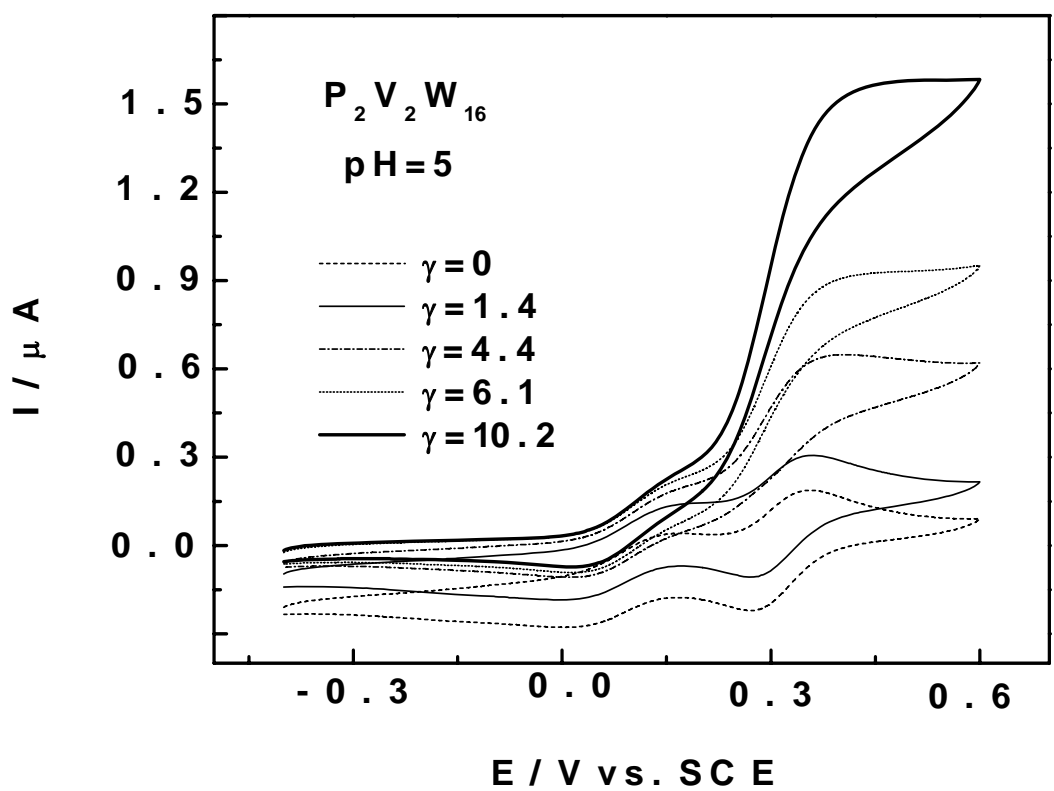

**Figure 2**

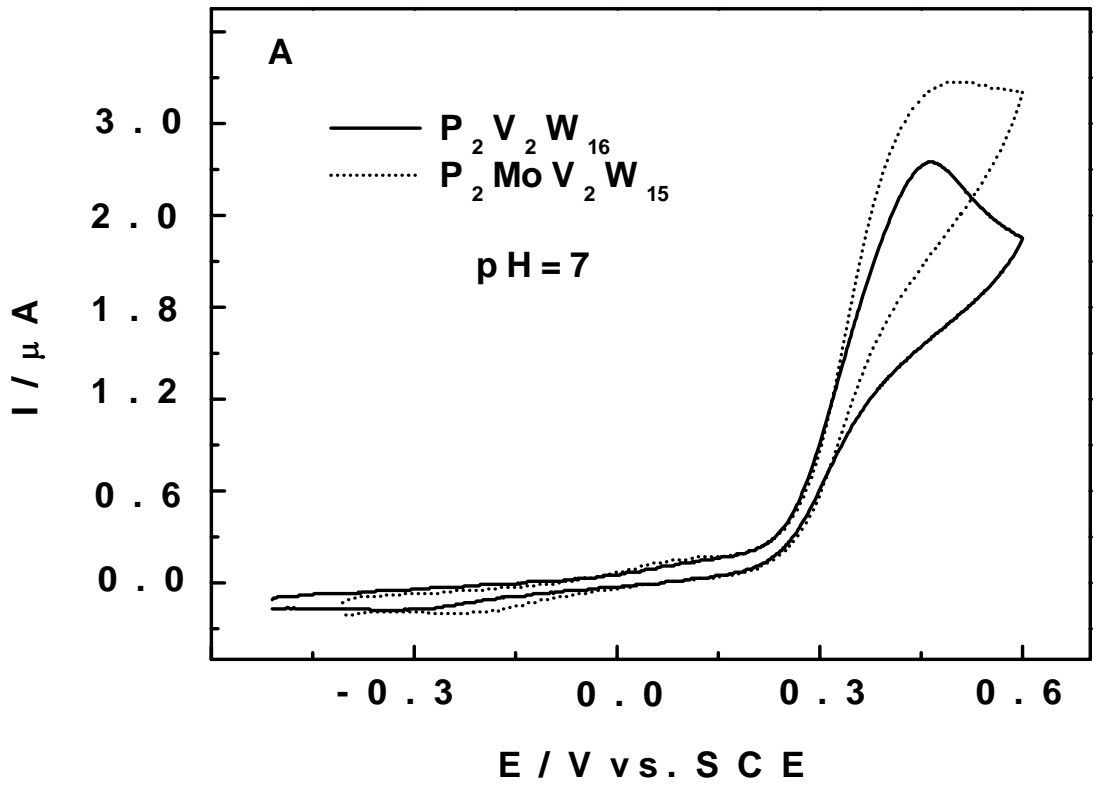

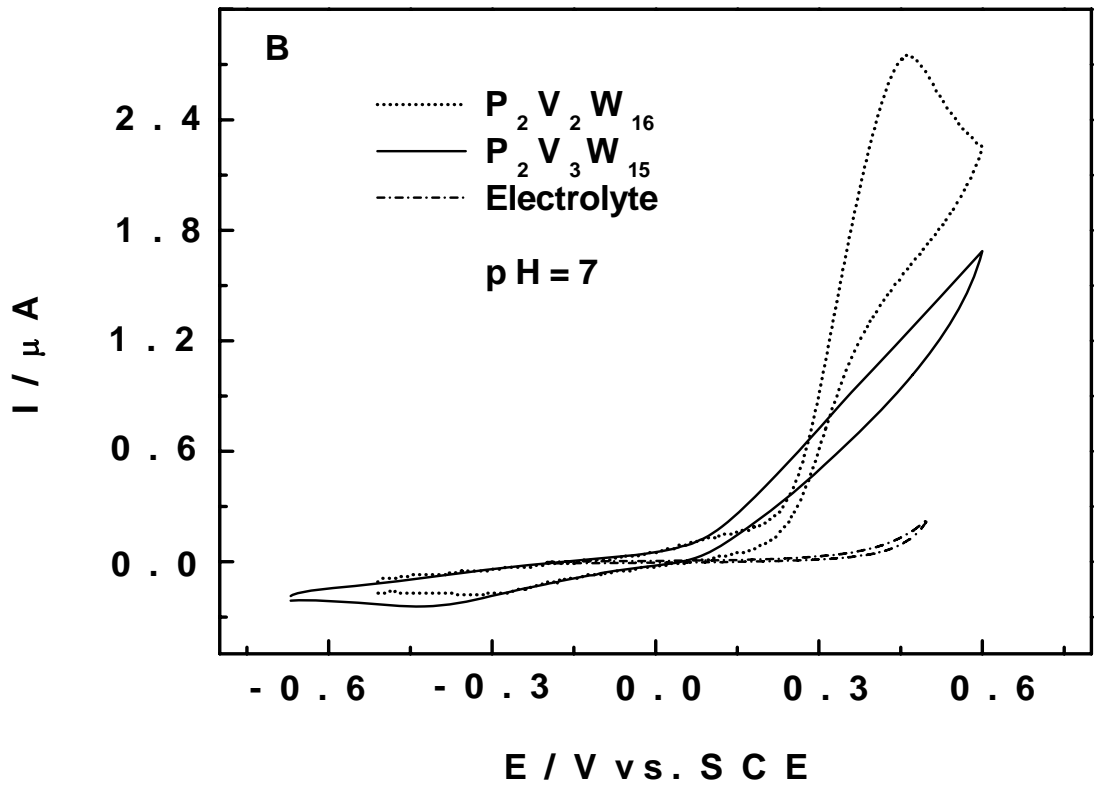

**Figure 3**

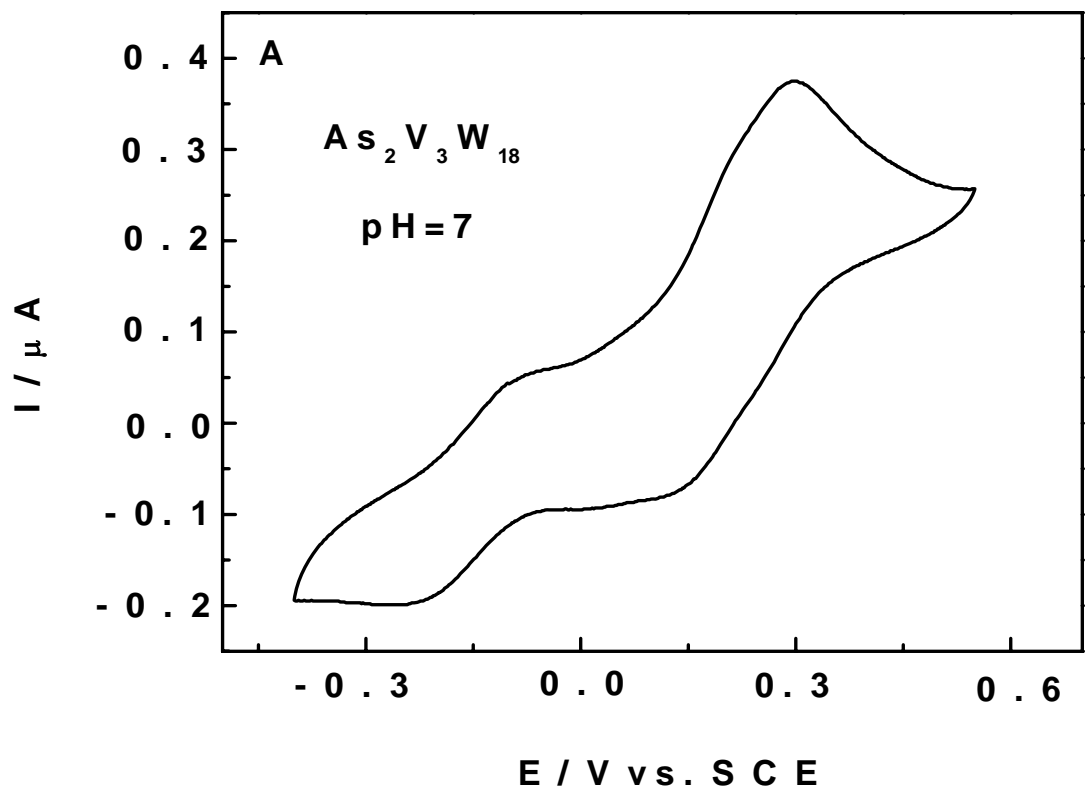

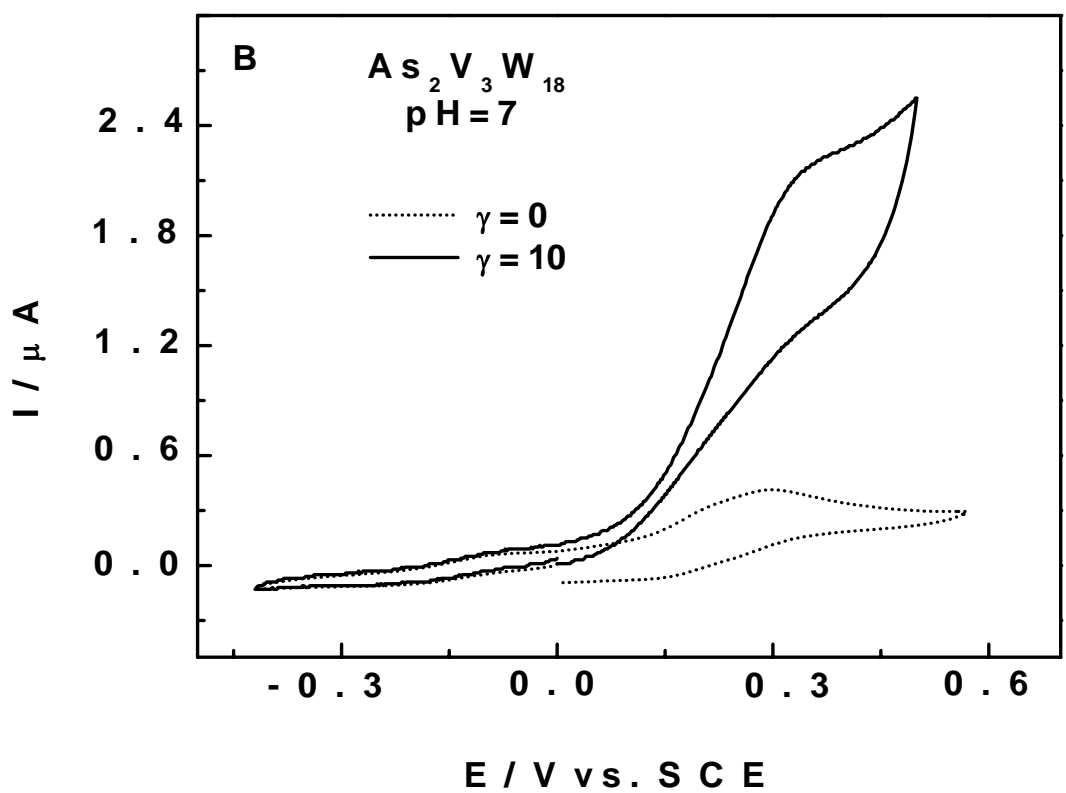

**Figure 4**